\documentclass[twocolumn]{aastex63}
\usepackage{apjfonts}

\received{2020 July 16}
\revised{2020 July 31}
\accepted{2020 August 12}
\shorttitle{Shock--Cloud Interaction in RX~J1713.7$-$3946}
\shortauthors{Tanaka et al.}

\begin{document}

\title{Shock--Cloud Interaction in the Southwestern Rim of RX~J1713.7$-$3946 Evidenced by Chandra X-ray Observations}

\correspondingauthor{Takaaki Tanaka}
\email{ttanaka@cr.scphys.kyoto-u.ac.jp}

\author[0000-0002-4383-0368]{Takaaki Tanaka}
\affiliation{Department of Physics, Kyoto University, Kitashirakawa Oiwake-cho, Sakyo, Kyoto 606-8502, Japan}

\author[0000-0003-1518-2188]{Hiroyuki Uchida}
\affiliation{Department of Physics, Kyoto University, Kitashirakawa Oiwake-cho, Sakyo, Kyoto 606-8502, Japan}

\author[0000-0003-2062-5692]{Hidetoshi Sano}
\affiliation{National Astronomical Observatory of Japan, 2-21-1 Osawa, Mitaka, Tokyo 181-8588, Japan}

\author[0000-0002-5504-4903]{Takeshi Go Tsuru}
\affiliation{Department of Physics, Kyoto University, Kitashirakawa Oiwake-cho, Sakyo, Kyoto 606-8502, Japan}



\begin{abstract}
We report on results of Chandra X-ray observations of the southwestern part of the supernova remnant (SNR) RX~J1713.7$-$3946. 
We measure proper motions of two X-ray bright blobs, named Blobs~A and B, in regions presumably corresponding to the forward shock of the SNR. 
The measured velocities are $3800 \pm 100~\mathrm{km}~\mathrm{s}^{-1}$ and  $2300 \pm 200~\mathrm{km}~\mathrm{s}^{-1}$ for Blobs~A and B, respectively. 
Since a dense molecular clump is located close to Blob~B, its slower velocity is attributed to shock deceleration as a result of a shock--cloud interaction. 
This result provides solid evidence that the forward shock of RX~J1713.7$-$3946 is indeed colliding with dense gas discovered through radio observations 
reported in the literature. 
The locations and velocity differences of the two blobs lead to an estimate that the shock encountered with the dense gas $\sim 100~\mathrm{yr}$ ago. 
The shock velocities, together with cutoff energies of the synchrotron X-ray spectra of the blobs, indicate that particle acceleration in these regions is  
close to the Bohm limit. Blob~B, in particular, is almost at the limit, accelerating particles at the fastest possible rate. 
We discuss possible influence of the shock--cloud interaction on the efficiency of particle acceleration. 
\end{abstract}



\section{Introduction} \label{sec:intro}
Supernova remnants (SNRs) have been attracting attention as one of the promising candidates for accelerators of Galactic cosmic rays \citep[e.g.,][]{Berezhko2014}. 
Nonthermal emissions in the X-ray and gamma-ray domains have been serving as observational probes of particles accelerated in expanding shocks of SNRs. 
Nonthermal X-rays detected in SNRs are almost exclusively attributed to synchrotron radiation 
from $\gtrsim \mathrm{TeV}$ electrons \citep[e.g.,][]{Koyama1995} 
except for a few exceptions claimed as nonthermal bremsstrahlung from sub-relativistic particles \citep[e.g.,][]{Tanaka2018}. 
Gamma-rays from a handful of SNRs are firmly confirmed as emission due to decay of $\pi^0$ mesons produced by interactions between accelerated protons and 
ambient gas \citep[e.g.,][]{Giuliani2011,Ackermann2013}. 
However, gamma-ray emissions detected in SNRs, including the target of the present work, RX~J1713.7$-$3946, 
can generally be explained also by inverse Compton scattering or bremsstrahlung from accelerated electrons, 
which makes their emission mechanisms still controversial \citep[e.g.,][]{Tanaka2011}. 

RX~J1713.7$-$3946 has been regarded as one of the most important SNRs for studies on particle acceleration because of its bright X-ray and gamma-ray 
nonthermal radiation. 
The X-ray emission is dominated by synchrotron radiation \citep[e.g.,][]{Koyama1997,Tanaka2008,Acero2009,Okuno2018} with barely detected thermal emission ascribed to reverse-shocked ejecta \citep{Katsuda2015}. 
The gamma-ray emission is detected in the GeV range with the Large Area Telescope onboard the Fermi Gamma-ray Space Telescope \citep{Abdo2011,HESS2018} 
and in the TeV range with the High Energy Stereoscopic System (H.E.S.S.) \citep{Aharonian2004,Aharonian2006,Aharonian2007,HESS2018}. 
The SNR is often associated with the Chinese ``guest star'' in AD~393 \citep{Wang1997} although the association is questioned by \cite{Fesen2012}. 
From the X-ray expansion measurements of the northwestern (NW) and southeastern (SE) rims, 
the age is independently estimated to be $\sim 1500\textrm{--}2300~\mathrm{yr}$, which is roughly consistent with the supernova explosion in AD~393 \citep{Tsuji2016,Acero2017}. 
The mostly accepted distance to RX~J1713.7$-$3946 is $1~\mathrm{kpc}$, deduced based on the X-ray absorption column density measured 
by \cite{Koyama1997} and on the distance to the molecular cloud associated with the SNR discovered by \cite{Fukui2003} in CO line data. 

\cite{Fukui2012} claimed that the shock wave of RX~J1713.7$-$3946 recently collided with a inhomogeneous dense gas 
wall created by stellar wind from the progenitor. 
Such dense gas serves as targets for accelerated protons in production of $\pi^0$ mesons. 
The TeV gamma-ray distribution in fact traces well that of molecular and atomic gas as reported by \cite{Fukui2012}. 
Shock--cloud interaction can play another role if the cloud is not uniform but clumpy. 
As revealed by \cite{Inoue2012} using magnetohydrodynamic (MHD) simulations, interactions with clumpy gas deform the shock front and 
leave turbulent eddies behind the shock, resulting in magnetic field amplification up to $0.1\textrm{--}1~\mathrm{mG}$. 
The amplified magnetic field makes the timescales for particle acceleration and synchrotron cooling shorter, probably causing 
the short-timescale variability of synchrotron X-rays found by \cite{Uchiyama2007}. 
The amplified magnetic field also enhances synchrotron radiation around the clump. 
\cite{Sano2013} indeed showed such synchrotron X-ray enhancement, supporting the prediction by \cite{Inoue2012}. 

We here report on results from Chandra X-ray observations of the southwestern (SW) rim of the SNR RX~J1713.7$-$3946, 
where dense molecular clumps are located \citep{Fukui2012}. 
We perform expansion measurements of the SNR shell using Chandra data taken in 2005 and 2020 
in order to obtain a clear signatures of a shock--cloud interaction. 
Performing spectral analysis, we then discuss the effect of the interaction on particle acceleration in terms of acceleration efficiency.   
Quoted uncertainties indicate $1\sigma$ confidence intervals throughout the Letter.

\begin{figure*}[bth]
\epsscale{1.18}
\plotone{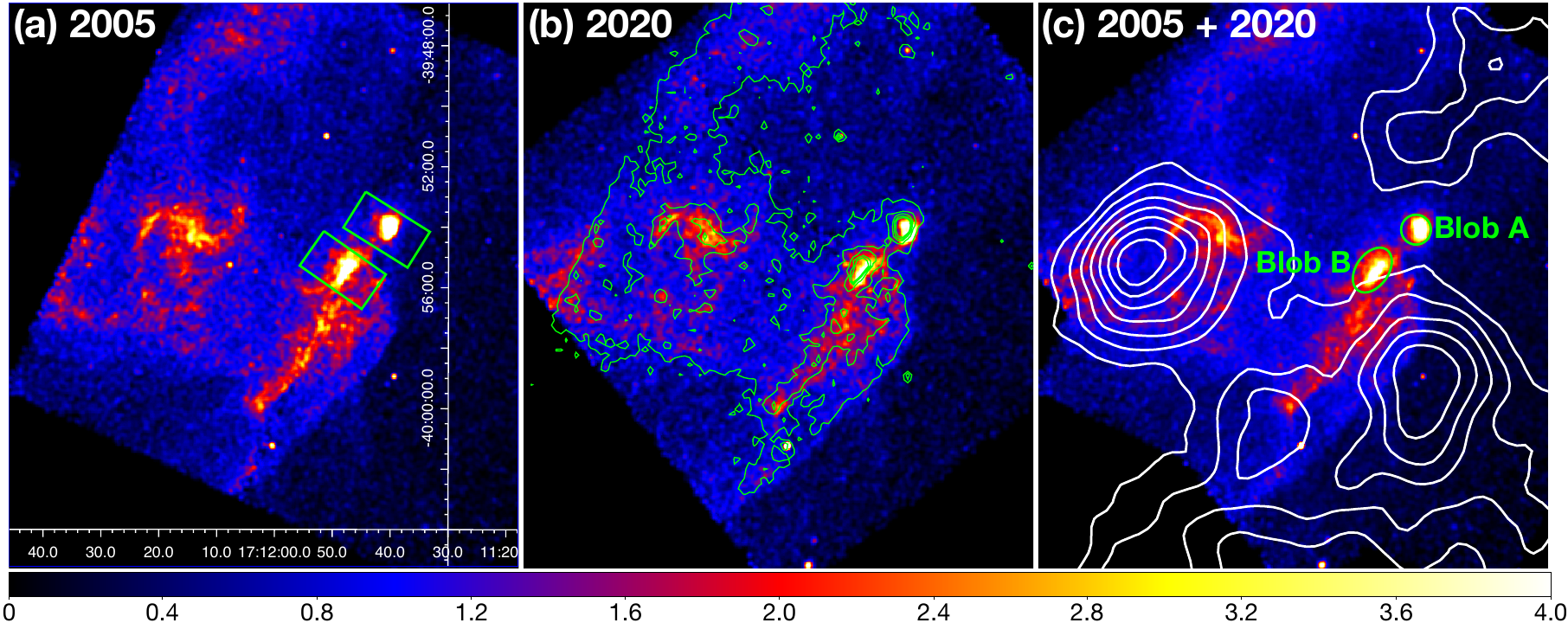}
\caption{
\label{fig:image}
(a) Exposure-corrected X-ray (0.7--7~keV) image obtained with Chandra ACIS from the observation in 2005. North is up, and east is to the left. The green rectangles are regions used for extracting profiles in presented in Figure~\ref{fig:profile}(a) and (b). The color scale indicates flux in a unit of $10^{-7}~\mathrm{ph}~\mathrm{cm}^{-2}~\mathrm{s}^{-1}$. The pixel size of the image is $2\arcsec \times 2\arcsec$. The image is smoothed with a Gaussian kernel with $\sigma = 4\arcsec$. (b) Same as (a) but from the observation in 2020. The image from 2005 is overlaid as the green contours to visualize proper motions. (c) Sum of the images from 2005 and 2020. The green circle and ellipse are the regions used for extracting the spectra of the two blobs. 
The white contours indicate the distribution of the $^{12}\mathrm{CO}\,(J=2\textrm{--}1)$ line emission as observed with NANTEN2 integrated over a velocity range from 
$-20.2~\mathrm{km}~\mathrm{s}^{-1}$ to $-0.2~\mathrm{km}~\mathrm{s}^{-1}$, where dense gas associated with RX~J1713.7$-$3946 is located \citep{Sano2013}. 
Each contour is drawn at every $5~\mathrm{K}~\mathrm{km}~\mathrm{s}^{-1}$ between $10$ and $40~\mathrm{K}~\mathrm{km}~\mathrm{s}^{-1}$. 
}
\end{figure*}

\section{Observations and Data Reduction} \label{sec:obs}
The first Chandra observation of the SW part of RX~J1713.7$-$3946 was performed in 2005 July (Obs ID: 5561) with ACIS-I. 
We performed another Chandra observation of almost the same region in 2020 May after a time interval of $\sim 15~\mathrm{yr}$ again with ACIS-I (Obs ID: 21339). 
Examining the light curves, we found no significant background flares during both observations. 
The effective exposure times are 29.0~ks and 29.7~ks for the observations in 2005 and 2020, respectively. 
We reprocessed the data using the Chandra Interactive Analysis of Observations (CIAO) version 4.12 and Chandra Calibration Database (CALDB) version 4.9.1. 

We aligned the data taken in 2020 to the coordinate of the data from the 2005 observation to make our  expansion measurements as accurate as possible. 
We detected point sources in the field-of-views of the observations with the {\tt wavdetect} tool in CIAO. 
Cross-matching nine sources detected, we computed a transformation matrix describing translation, rotation, and scaling with {\tt wcs\_match} in CIAO. 
We then reprojected the events file from 2005 using {\tt wcs\_update}.

\begin{figure*}
\epsscale{1.1}
\plotone{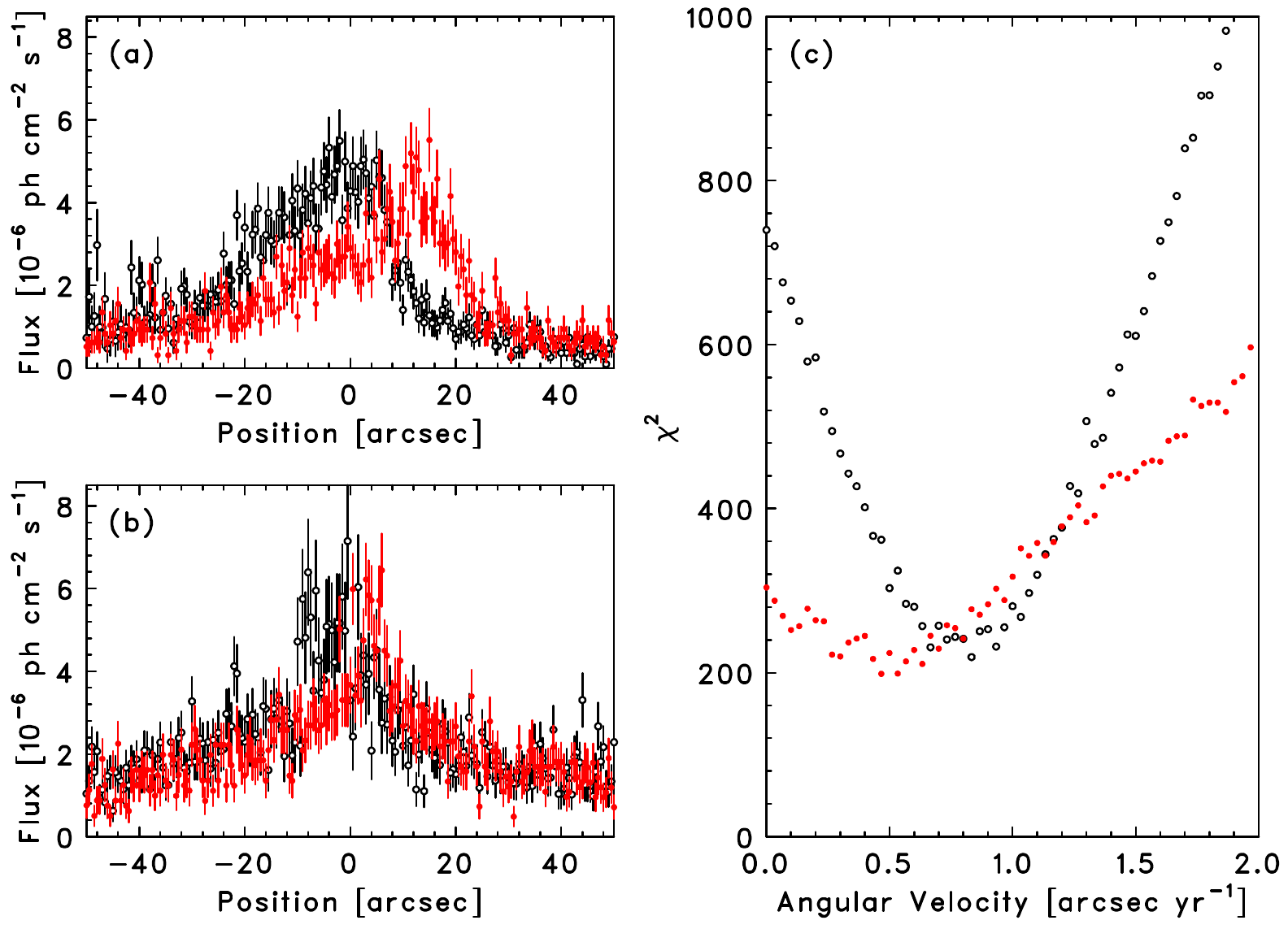}
\caption{
\label{fig:profile}
(a) Profile along the rectangular region containing Blob~A in Figure~\ref{fig:image}(a). The shock upstream is to the right. The black and red points indicate data from 2005 and 2020, respectively. The origin of the horizontal axis is set so that the peak of the profile in 2005 comes to $\sim 0$. 
(b) Same as (a) but for Blob~B. 
(c) $\chi^2$ as a function of angular velocity for Blobs~A (black) and B (red). 
}
\end{figure*}

\section{Analysis and Results} \label{sec:ana}
Figure~\ref{fig:image} presents 0.7--7~keV exposure-corrected images of the SW region of RX~J1713.7$-$3946 as observed with Chandra ACIS-I in 2005 and 2020. 
Also shown is the sum of the two images. 
Comparing the two images from 2005 and 2020, one can clearly see the expansion of the shell during the 15-yr time interval. 
As illustrated in Figure~\ref{fig:image}(c), a molecular clump is detected through CO, CS, HC$_3$N, and SiO line observations just outside the western edge \citep{Fukui2003,Moriguchi2005,Sano2010,Sano2013,Sano2015,Fukui2012,Maxted2012}.  
If the shock is actually interacting with the molecular gas, the shock velocity would be substantially lower than those 
measured in other locations of the SNR by \cite{Tsuji2016} and \cite{Acero2017}. 

In what follows, we focus on the two bright blobs, Blobs~A and B (Figure~\ref{fig:image}(c)), at the western edge of the shell. 
In Figure~\ref{fig:profile}(a) and (b), we plot projected profiles of the blobs along the regions shown in Figure~\ref{fig:image}(a). 
The rotation angles of the regions were selected so that they roughly accord with the directions of the proper motions. 
The expansion is again visible for both blobs. 
Blob B appears to have a lower shock velocity than Blob A. 
We quantified the velocities by comparing the profiles obtained in the two epochs. 
We artificially shifted the profile in 2020 and searched for a shift that gives the best match with the profile in 2005 in terms of $\chi^2$ defined as 
\begin{eqnarray}
\chi^2 =  \sum_i \frac{(f_i - g_i)^2}{(df_i)^2 + (dg_i)^2}, 
\end{eqnarray}
where $i$ in the index for bins of the profile histograms, $f_i$ and $g_i$ are fluxes in bin $i$, and $df_i$ and $dg_i$ 
are their statistical errors. 
We did not limit the shift to an integer multiple of the bin width, $0.5\arcsec$. 
We rebinned the shifted profile with the same bins as the histogram for the profile in 2005 assuming that the profile is uniform inside each bin. 
Figure~\ref{fig:profile}(c) is the $\chi^2$ profiles obtained for the blobs. 
We fitted it with a quadratic function and obtained a velocity that gives the minimum $\chi^2$ $(= {\chi_\mathrm{min}}^2)$. 
A velocity range that satisfies $\chi^2 \leq {\chi_\mathrm{min}}^2 +1$ is quoted as a $1\sigma$ confidence region. 
The resultant angular velocities are $0\farcs81\pm0\farcs03~\mathrm{yr}^{-1}$ and $0\farcs49\pm0\farcs05~\mathrm{yr}^{-1}$ 
for Blobs~A and B, respectively. 
If the distance to RX~J1713.7$-$3946 is $1~\mathrm{kpc}$, they are translated to velocities of $3800 \pm 100~\mathrm{km}~\mathrm{s}^{-1}$ 
and $2300 \pm 200~\mathrm{km}~\mathrm{s}^{-1}$. 
The velocity of Blob~B is indeed significantly slower than that of Blob~A.

\begin{figure}
\epsscale{1.15}
\plotone{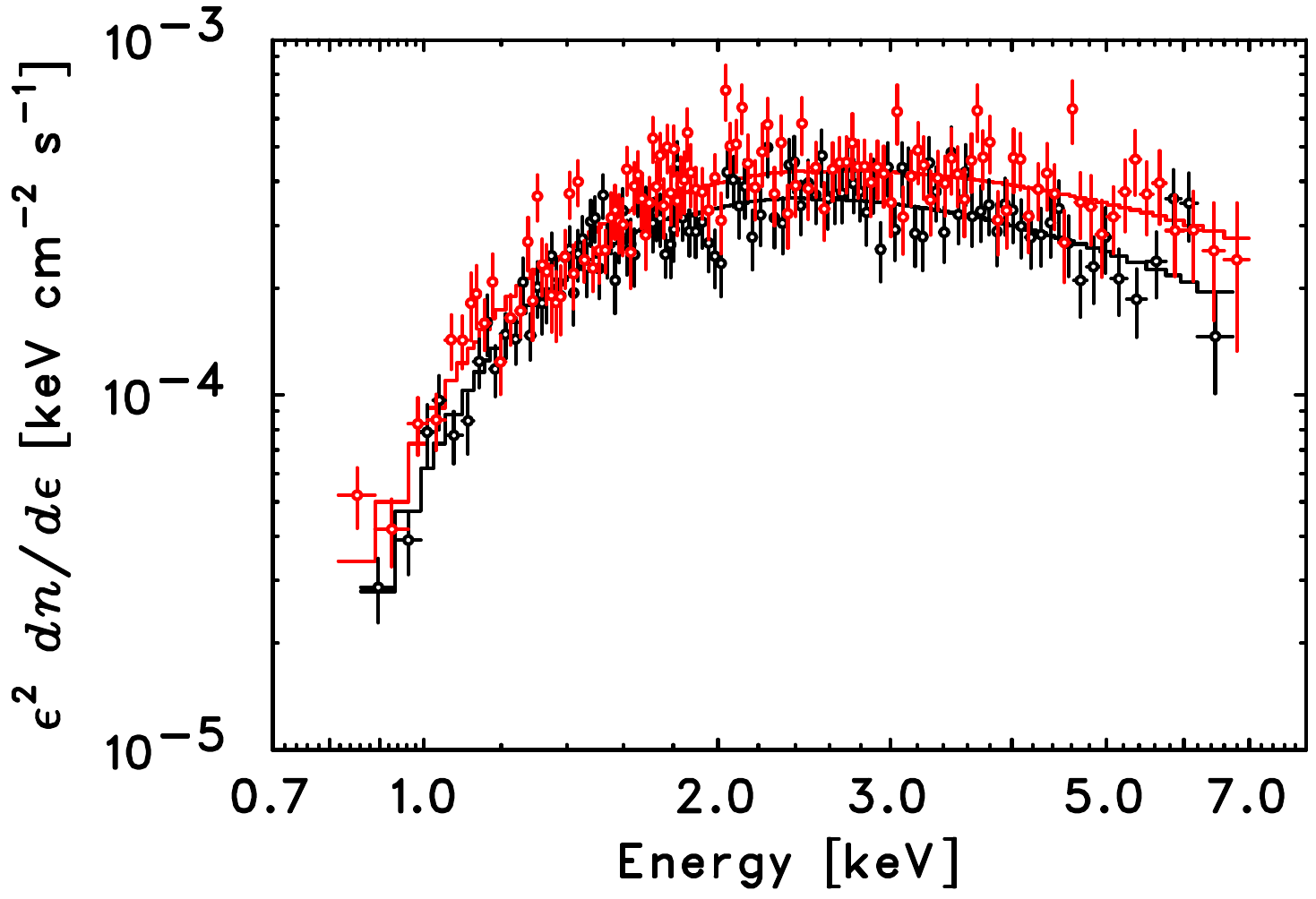}
\caption{
\label{fig:spec}
Unfolded spectra of Blobs~A (black points) and B (red points). 
The solid lines indicate the best-fit ZA07 models (see the text for details). 
}
\end{figure}

We also analyzed spectra of the two blobs extracted from the regions indicated in Figure~\ref{fig:image}(c). 
Since we did not see any significant differences between the spectra from 2005 and 2020, we combined those from the two epochs for each blob. 
The background spectra were extracted from off-source regions in the same field-of-views. 
We plot the background-subtracted spectra in Figure~\ref{fig:spec}. 
In the following spectral fittings, we used {\tt XSPEC} version 12.10.0 \citep{Arnoud1996} with the solar abundance table 
based on the result by \cite{Wilms2000}. 
We modeled the interstellar absorption with the Tuebingen-Boulder model ({\tt TBabs}; \citealt{Wilms2000}) implemented in {\tt XSPEC}. 
The minimum $\chi^2$ statistic was used for the spectral fittings. The spectra were binned so that each bin has at least 30 counts.   

We first fitted the spectra with a phenomenological model, a power law modified by interstellar absorption, which we refer to as PL. 
Table~\ref{tab:spec} summarizes the results, which agree well with those by \cite{Okuno2018} analyzing the data from 2005. 
We tried another set of spectral fittings by replacing the power law with a more physically oriented model taken from \cite{Zirakashvili2007}, who 
gave analytical descriptions of synchrotron radiation spectra under the assumption that electron energy losses are dominated by synchrotron cooling. 
We adopt here their formula for synchrotron spectra produced downstream of the shock with the downstream magnetic field stronger than upstream 
by a factor of $\kappa^{-1} = \sqrt{11}$, namely, 
\begin{eqnarray} \label{eq:ZA}
\frac{dn}{d\varepsilon} \propto \left( \frac{\varepsilon}{\varepsilon_0}  \right)^{-2} \left( 1 + 0.38 \sqrt{\frac{\varepsilon}{\varepsilon_0}} \right)^{11/4} \exp \left( -\sqrt{\frac{\varepsilon}{\varepsilon_0}} \right), 
\end{eqnarray}
where $\varepsilon$ denotes the photon energy and $\varepsilon_0$ is the cutoff energy. 
Note that the actual cutoff is located at $\varepsilon \geq 10\, \varepsilon_0$ as pointed out by \cite{Zirakashvili2007}. 
The results with this model are again summarized in Table~\ref{tab:spec}, where we call the model ZA07. 
Blob~B has a higher cutoff energy as is expected from its harder spectrum. 

\begin{deluxetable*}{cccccccc}[bt]
\tablenum{1}
\tablecaption{\label{tab:spec}Best-fit spectral parameters.}
\tablewidth{0pt}
\tablehead{
\colhead{Region} & \colhead{Model} & \colhead{$N_\mathrm{H}$\tablenotemark{a}} & \colhead{$\Gamma$\tablenotemark{b}} & \colhead{$\varepsilon_0$} & \colhead{$F_{1-5~\mathrm{keV}}$\tablenotemark{c}}& \colhead{$\chi^2$} &  \colhead{d.o.f} \\
 & & ($10^{22}~\mathrm{cm}^{-2}$) & & (keV) & ($10^{-12}~\mathrm{erg}~\mathrm{cm}^{-2}~\mathrm{s}^{-1}$) & &
}
\startdata
Blob~A & PL & $1.58^{+0.09}_{-0.08}$  & $2.74 \pm 0.07$ & \nodata & $1.34 \pm 0.06$ & 109.7 & 109\\
Blob~B & PL & $1.38 \pm 0.09$ & $2.52 \pm 0.08$ & \nodata & $1.47^{+0.07}_{-0.06}$ & 141.0 & 117 \\
\hline
Blob~A & ZA07 & $1.38 \pm 0.07$ & \nodata & $0.40^{+0.06}_{-0.05}$ & $1.21 \pm 0.04$ & 107.7 & 109\\
Blob~B & ZA07 & $1.23 \pm 0.07$ & \nodata & $0.62^{+0.13}_{-0.10} $ & $1.38 \pm 0.05$ & 138.7 & 117 \\
\enddata
\tablenotetext{a}{Equivalent hydrogen column density.}
\tablenotetext{b}{Photon index.}
\tablenotetext{b}{Unabsorbed flux integrated from $1~\mathrm{keV}$ to $5~\mathrm{keV}$.}
\end{deluxetable*}

\section{Discussion} \label{sec:dis}
We measured the proper motions of the two bright blobs in the SW rim of RX~J1713.7$-$3946. 
The proper motion velocities cannot necessarily be regarded as shock velocities because of possible line-of-sight velocities. 
Blob~A is at the very edge of the SNR, and thus the measured velocity of $3800~\mathrm{km}~\mathrm{s}^{-1}$ would safely be taken 
as the shock velocity ($V_\mathrm{sh}$) of the region. 
Blob~B, on the other hand, appears located $\lesssim 2\arcmin$ inward from the edge of the faint diffuse emission extending 
beyond the bright structures (Figure~\ref{fig:image}). 
Given the radius of the remnant , $\sim 30\arcmin$ (see, e.g., \citealt{Acero2009} for an X-ray image of the whole SNR), 
we can infer that the actual shock velocity of Blob~B would be a factor of $\lesssim 1.07$ larger than the proper motion, 
or $V_\mathrm{sh} \lesssim 2500~\mathrm{km}~\mathrm{s}^{-1}$, assuming a spherical expansion. 
Since this projection effect is small and the relevant assumptions should have some uncertainties, we 
treat the measured velocity as $V_\mathrm{sh}$ also for Blob~B. 
Note that this does not substantially affect the following discussion. 

The two regions we studied have significantly different shock velocities in spite of their proximity. 
A clue to understanding it can be found in the distribution of the interstellar gas shown in Figure~\ref{fig:image}(c). 
Blob~A has no noticeable cloud nearby while the shell structure including Blob~B seems in contact with the molecular clump. 
A plausible explanation of the velocity difference is that Blob~B collided with the clump at some point of the SNR evolution and 
was decelerated. 
Further supporting this scenario is the morphology in which Blob~B is trailing Blob~A. 
Then, the present work has provided the most direct evidence ever that the shock of RX~J1713.7$-$3946 is indeed interacting with 
the cloud found in radio observations \citep{Fukui2003,Moriguchi2005,Sano2010,Sano2013,Sano2015,Fukui2012,Maxted2012}. 

From the shock velocities, we can estimate when the shock started to interact with the molecular cloud. 
Blob~A was running ahead of Blob~B by $\sim 0\farcm5$, corresponding to $0.15~\mathrm{pc}$ at a distance of $1~\mathrm{kpc}$, 
at the point of the observation in 2020 (Figure~\ref{fig:image}). 
Since the velocity difference between the two blobs is $1500~\mathrm{km}~\mathrm{s}^{-1} = 1.5 \times 10^{-3}~\mathrm{pc}~\mathrm{yr}^{-1}$, 
the two blobs were located at the same radius of the SNR $\sim 100~\mathrm{yr}$ under an assumption 
that the velocities have been almost constant during the time interval. 
This would give the first-order estimate of the epoch when the shock in the Blob-B region encountered the molecular clump. 
Our estimate is roughly consistent with the suggestion by \cite{Fukui2012} that the forward shock of RX~J1713.7$-$3946 expanded almost 
freely in a cavity in the early phase of its evolution and collided with a dense gas wall swept up by the stellar wind from the progenitor 
a few $100~\mathrm{yr}$ ago. 

When the shock collided with the gas wall $\sim 100~\mathrm{yr}$ ago, the distance between the central compact object 1WGA~J1713.4$-$3946 
and the blobs would have been $\sim 6~\mathrm{pc}$ (at a distance of 1~kpc), which would correspond to the radius of the wind-blown bubble. 
According to \cite{Chevalier1999}, a star with a main-sequence mass of $\sim 15\,M_\sun$ is capable of creating a bubble with 
such a radius. The mass agrees well with the estimate based on elemental abundances of the supernova ejecta by \cite{Katsuda2015}.
From an interacting shock, one may expect forbidden lines from low ionized ions in the optical and infrared bands. 
However, these lines become bright only after the shock enters the radiative phase \citep{Lee2015}, and thus we cannot expect them in the present case where 
the shock collided with dense gas recently.

It would be worth pointing out that the velocity of Blob~A ($V_\mathrm{sh} = 3800~\mathrm{km}~\mathrm{s}^{-1}$),
which does not seem to be interacting with dense gas, is comparable to $V_\mathrm{sh} = 3900~\mathrm{km}~\mathrm{s}^{-1}$ 
in the NW reported by \cite{Tsuji2016} and also to $V_\mathrm{sh} = 3500~\mathrm{km}~\mathrm{s}^{-1}$ in the SE measured by \cite{Acero2017}. 
Using the shock velocities in the NW and SE, the authors performed hydrodynamical analysis of the SNR evolution and reached conclusions  
that the age of the SNR is within a range of $\sim 1500\textrm{--}2000~\mathrm{yr}$. 
The present result supports their conclusions about the age with a similar velocity measured for Blob~A. 
A question here is why only the shock velocity in the Blob~B is significantly decelerated although the shocks in NW and SE are suggested to 
be interacting with dense gas as well \citep[e.g.,][]{Fukui2012,Sano2013,Sano2015}. 
One of the possible answers would be that the shock in the Blob-B region is interacting with denser gas than the shocks in other regions. 
The gas distribution map by \cite{Fukui2012} in fact indicates a higher gas density in the SW, supporting our hypothesis.

\begin{figure}[htb]
\epsscale{1.15}
\plotone{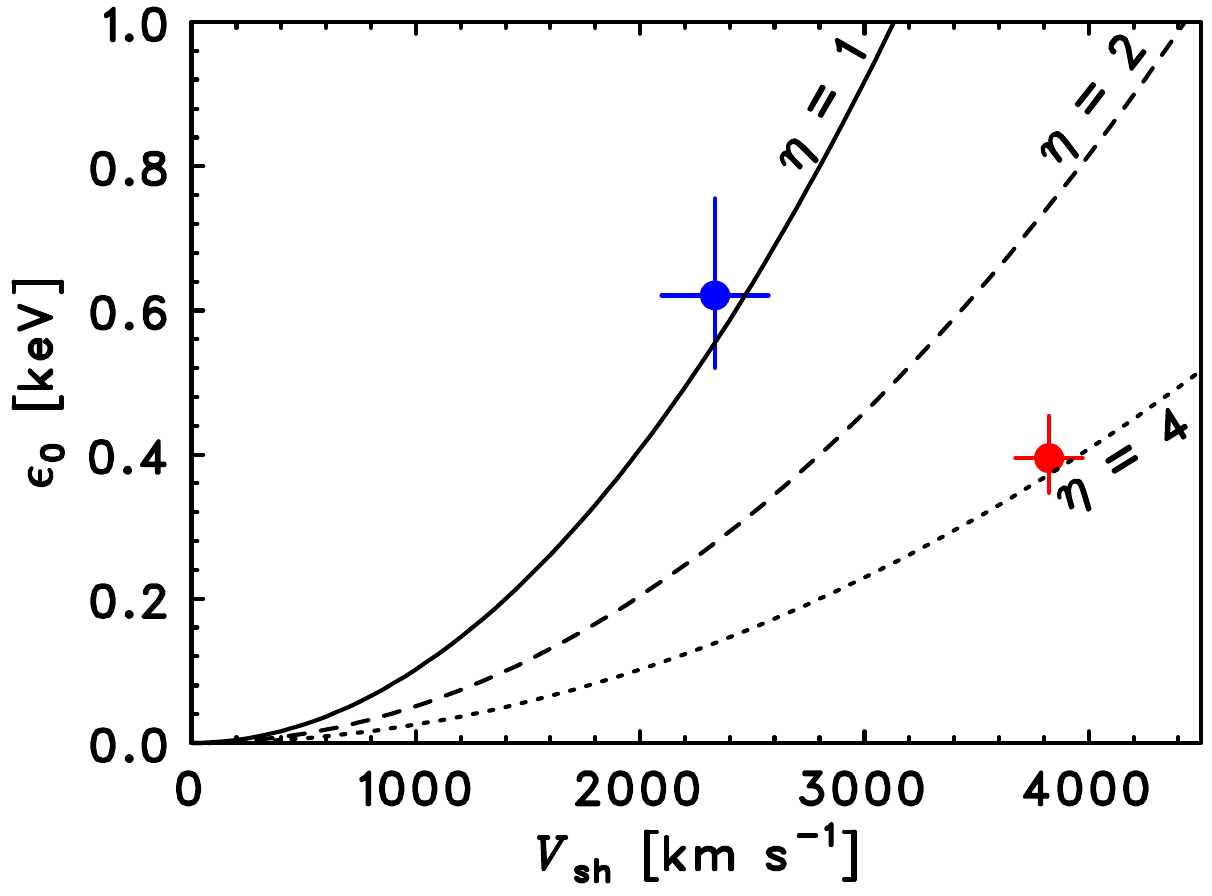}
\caption{
\label{fig:eps0-Vsh} 
Cutoff energy ($\varepsilon_0$) from the spectral fittings with the model by \cite{Zirakashvili2007} (Equation~(\ref{eq:ZA})) plotted against the measured shock velocity ($V_\mathrm{sh}$). 
The red and blue points are from Blobs~A and B, respectively. 
The black curves represent the relations predicted by Equation~(\ref{eq:cutoff}) for $\eta = 1$ (solid), 2 (dashed), and 4 (dotted). 
}
\end{figure}

An interesting finding in our spectral analysis is that Blob~B has a harder spectrum than Blob~A despite the slower shock velocity. 
The X-ray band corresponds to the cutoff region of the synchrotron spectrum. 
Thus, a harder spectrum implies a higher cutoff energy as demonstrated in the results from the spectral fittings with the ZA07 model (Table~\ref{tab:spec}). 
When the electron maximum energy is limited by synchrotron cooling, 
the cutoff appears at an energy where the synchrotron cooling timescale is equal to the acceleration timescale. 
According to \cite{Zirakashvili2007}, the synchrotron cutoff energy is expressed as
\begin{eqnarray} \label{eq:cutoff}
\varepsilon_0 = 0.92 \left( \frac{V_\mathrm{sh}}{3000~\mathrm{km}~\mathrm{s}^{-1}} \right)^2  \eta^{-1}~\mathrm{keV}, 
\end{eqnarray}
where $\eta~(\geq 1)$ is the gyrofactor. 
The equation is for $\kappa^{-1} = 1/\sqrt{11}$, the same as Equation~(\ref{eq:ZA}). 
We can compute $\eta$ in the two blobs by substituting $V_\mathrm{sh}$ and $\varepsilon_0$ of Equation~(\ref{eq:cutoff}) 
with the values obtained in \S\ref{sec:ana}. 
We obtain $\eta = 3.8^{+0.5}_{-0.6}$ and $\eta = 0.9^{+0.2}_{-0.3}$ for Blobs~A and B, respectively. 
In Figure~\ref{fig:eps0-Vsh}, we plot $\varepsilon_0$ against $V_\mathrm{sh}$ together with equi-$\eta$ curves. 
Particle acceleration in both blobs is close to the Bohm limit ($\eta = 1$), similarly to the results based on spectra extracted 
from much larger regions \citep{Uchiyama2007,Tanaka2008} and from another location of the SNR, NW \citep{Tsuji2019}. 
Blob~B has a gyrofactor almost at the  limit, with which particle acceleration proceeds at the fastest possible rate.

It is often supposed that a higher shock velocity makes diffusive shock acceleration more efficient. 
On the contrary, our result implies that Blob~B, which is interacting with the dense molecular clump and has a lower shock velocity, 
has a smaller $\eta$ and thus is accelerating particles more efficiently. 
The gyrofactor $\eta$ is related to the turbulence in the magnetic field as $\eta = (B/\delta B)^2$, where $B$ is the strength of the static 
magnetic field and $\delta B$ is the turbulence level. 
It seems that magnetic turbulence is somehow induced by the shock--cloud interaction in Blob~B. 
The MHD simulations by \cite{Inoue2012} indicate that, when a shock interacts with clumpy gas,  turbulence is generated and 
the magnetic field is amplified around clumps. 
This is also supported by the results from MHD simulations by \cite{Celli2019}. 
\cite{Sano2013} and \cite{Sano2015} compared distribution of synchrotron X-rays of RX~J1713.7$-$3946 with gas distribution and 
claimed that the mechanism proposed by  \cite{Inoue2012} is at work in this SNR. 
The same mechanism may be able to explain the efficient particle acceleration in Blob~B.

Although we so far assumed implicitly that particles are accelerated at the forward shock, it would be possible that particles are reaccelerated 
at reflected shocks generated due to the shock--cloud interaction. 
In the downstream region where reflected shocks propagate, magnetic turbulence is enhanced by the mechanism mentioned above. 
Therefore, fast particle acceleration is possible in reflected shocks \citep{Inoue2012}. 
It would be possible that at least a part of the synchrotron X-rays from Blob~B is emitted by electrons in reflected shocks. 
To further discuss the possibility of such a scenario, it would be essential to reveal the distribution of the gas in this region with superior 
angular resolution through observations by, e.g, the Atacama Large Millimeter/submillimeter Array. 

\acknowledgments
We thank staff at the Chandra X-ray Center (CXC) for supporting our observation. 
This research has made use of data obtained from the Chandra Data Archive and software provided by the CXC in the application package CIAO. 
NANTEN2 is an international collaboration of 10 universities: Nagoya University, Osaka Prefecture University, University of Cologne, University of Bonn, Seoul National University, University of Chile, University of New South Wales, Macquarie University, University of Sydney, and University of ETH Zurich. 
We appreciate Tomoyuki Okuno for his help in preparing the proposal for the Chandra observation in 2020. 
We also thank Hiroya Yamaguchi and Shiu-Hang Lee for fruitful discussions. 
This work is supported by JSPS Scientific Research grant Nos. JP19H01936 (T.T.), JP19K03915 (H.U.), JP19H05075 (H.S.), and JP19K14758 (H.S.).
%




\begin{thebibliography}{}
\bibitem[Abdo et al.(2011)]{Abdo2011} Abdo, A.~A., Ackermann, M., Ajello, M., et al.\ 2011, \apj, 734, 28

\bibitem[Acero et al.(2017)]{Acero2017} Acero, F., Katsuda, S., Ballet, J., et al.\ 2017, \aap, 597, A106

\bibitem[Acero et al.(2009)]{Acero2009} Acero, F., Ballet, J., Decourchelle, A., et al.\ 2009, \aap, 505, 157

\bibitem[Ackermann et al.(2013)]{Ackermann2013} Ackermann, M., Ajello, M., Allafort, A., et al.\ 2013, Science, 339, 807

\bibitem[Aharonian et al.(2007)]{Aharonian2007} Aharonian, F., Akhperjanian, A.~G., Bazer-Bachi, A.~R., et al.\ 2007, \aap, 464, 235

\bibitem[Aharonian et al.(2006)]{Aharonian2006} Aharonian, F., Akhperjanian, A.~G., Bazer-Bachi, A.~R., et al.\ 2006, \aap, 449, 223

\bibitem[Aharonian et al.(2004)]{Aharonian2004} Aharonian, F.~A., Akhperjanian, A.~G., Aye, K.-M., et al.\ 2004, \nat, 432, 75

\bibitem[Arnaud(1996)]{Arnoud1996} Arnaud, K.~A.\ 1996, adass V, 101, 17

\bibitem[Berezhko(2014)]{Berezhko2014} Berezhko, E.~G.\ 2014, Nuclear Physics B Proceedings Supplements, 256, 23

\bibitem[Celli et al.(2019)]{Celli2019} Celli, S., Morlino, G., Gabici, S., et al.\ 2019, \mnras, 487, 3199

\bibitem[Chevalier(1999)]{Chevalier1999} Chevalier, R.~A.\ 1999, \apj, 511, 798

\bibitem[Fesen et al.(2012)]{Fesen2012} Fesen, R.~A., Kremer, R., Patnaude, D., et al.\ 2012, \aj, 143, 27

\bibitem[Fukui et al.(2012)]{Fukui2012} Fukui, Y., Sano, H., Sato, J., et al.\ 2012, \apj, 746, 82

\bibitem[Fukui et al.(2003)]{Fukui2003} Fukui, Y., Moriguchi, Y., Tamura, K., et al.\ 2003, \pasj, 55, L61

\bibitem[Giuliani et al.(2011)]{Giuliani2011} Giuliani, A., Cardillo, M., Tavani, M., et al.\ 2011, \apjl, 742, L30

\bibitem[H.E.S.S. Collaboration et al.(2018)]{HESS2018} H.E.S.S. Collaboration, Abdalla, H., Abramowski, A., et al.\ 2018, \aap, 612, A6

\bibitem[Inoue et al.(2012)]{Inoue2012} Inoue, T., Yamazaki, R., Inutsuka, S., et al.\ 2012, \apj, 744, 71

\bibitem[Katsuda et al.(2015)]{Katsuda2015} Katsuda, S., Acero, F., Tominaga, N., et al.\ 2015, \apj, 814, 29 

\bibitem[Koyama et al.(1997)]{Koyama1997} Koyama, K., Kinugasa, K., Matsuzaki, K., et al.\ 1997, \pasj, 49, L7

\bibitem[Koyama et al.(1995)]{Koyama1995} Koyama, K., Petre, R., Gotthelf, E.~V., et al.\ 1995, \nat, 378, 255

\bibitem[Lee et al.(2015)]{Lee2015} Lee, S.-H., Patnaude, D.~J., Raymond, J.~C., et al.\ 2015, \apj, 806, 71

\bibitem[Maxted et al.(2012)]{Maxted2012} Maxted, N.~I., Rowell, G.~P., Dawson, B.~R., et al.\ 2012, \mnras, 422, 2230

\bibitem[Moriguchi et al.(2005)]{Moriguchi2005} Moriguchi, Y., Tamura, K., Tawara, Y., et al.\ 2005, \apj, 631, 947

\bibitem[Okuno et al.(2018)]{Okuno2018} Okuno, T., Tanaka, T., Uchida, H., et al.\ 2018, \pasj, 70, 77


\bibitem[Sano et al.(2015)]{Sano2015} Sano, H., Fukuda, T., Yoshiike, S., et al.\ 2015, \apj, 799, 175

\bibitem[Sano et al.(2013)]{Sano2013} Sano, H., Tanaka, T., Torii, K., et al.\ 2013, \apj, 778, 59

\bibitem[Sano et al.(2010)]{Sano2010} Sano, H., Sato, J., Horachi, H., et al.\ 2010, \apj, 724, 59

\bibitem[Tanaka et al.(2018)]{Tanaka2018} Tanaka, T., Yamaguchi, H., Wik, D.~R., et al.\ 2018, \apjl, 866, L26

\bibitem[Tanaka et al.(2011)]{Tanaka2011} Tanaka, T., Allafort, A., Ballet, J., et al.\ 2011, \apjl, 740, L51

\bibitem[Tanaka et al.(2008)]{Tanaka2008} Tanaka, T., Uchiyama, Y., Aharonian, F.~A., et al.\ 2008, \apj, 685, 988

\bibitem[Tsuji et al.(2019)]{Tsuji2019} Tsuji, N., Uchiyama, Y., Aharonian, F., et al.\ 2019, \apj, 877, 96

\bibitem[Tsuji \& Uchiyama(2016)]{Tsuji2016} Tsuji, N., \& Uchiyama, Y.\ 2016, \pasj, 68, 108

\bibitem[Uchiyama et al.(2007)]{Uchiyama2007} Uchiyama, Y., Aharonian, F.~A., Tanaka, T., et al.\ 2007, \nat, 449, 576

\bibitem[Wang et al.(1997)]{Wang1997} Wang, Z.~R., Qu, Q.-Y., \& Chen, Y.\ 1997, \aap, 318, L59

\bibitem[Wilms et al.(2000)]{Wilms2000} Wilms, J., Allen, A., \& McCray, R.\ 2000, \apj, 542, 914

\bibitem[Zirakashvili \& Aharonian(2007)]{Zirakashvili2007} Zirakashvili, V.~N., \& Aharonian, F.\ 2007, \aap, 465, 695


\end{thebibliography}
\end{document}